\documentclass[twocolumn]{aastex631}

\usepackage{url,amssymb,amsmath,color,units,wasysym,epsfig,epstopdf,enumerate,tabularx, subfigure,hyperref}
\usepackage{makecell}
\hypersetup{colorlinks=true, citecolor=blue, urlcolor=blue, linkcolor=blue}
\usepackage{graphicx}
\usepackage{dcolumn}
\usepackage{bm}
 
\usepackage{newtxtext,newtxmath}
\usepackage[T1]{fontenc}
\usepackage{ae,aecompl}

\begin{document}

\title{Nuclear physics with gravitational waves from neutron stars disrupted by black holes}

\correspondingauthor{Teagan Clarke}
\email{teagan.clarke@monash.edu}

\author[0000-0002-6714-5429]{Teagan A. Clarke}
\affiliation{School of Physics and Astronomy, Monash University, VIC 3800, Australia}
\affiliation{OzGrav: The ARC Centre of Excellence for Gravitational-wave Discovery, Clayton, VIC 3800, Australia}

\author{Lani Chastain}
\affiliation{Department of Physics \& Astronomy, University of North Georgia, GA 30597, USA}

\author[0000-0003-3763-1386]{Paul D. Lasky}
\affiliation{School of Physics and Astronomy, Monash University, VIC 3800, Australia}
\affiliation{OzGrav: The ARC Centre of Excellence for Gravitational-wave Discovery, Clayton, VIC 3800, Australia}

\author[0000-0002-4418-3895]{Eric Thrane}
\affiliation{School of Physics and Astronomy, Monash University, VIC 3800, Australia}
\affiliation{OzGrav: The ARC Centre of Excellence for Gravitational-wave Discovery, Clayton, VIC 3800, Australia}

\date{\today}

\begin{abstract}

Gravitational waves from neutron star-black hole (NSBH) mergers that undergo tidal disruption provide a potential avenue to study the equation of state of neutron stars and hence the behaviour of matter at its most extreme densities. 
We present a phenomenological model for the gravitational-wave signature of tidal disruption, which allows us to measure the disruption time. 
We carry out a study with mock data, assuming an optimistically nearby NSBH event with parameters tuned for measuring the tidal disruption. 
We show that a two-detector network of \unit[40]{km} Cosmic Explorer instruments can measure the time of disruption with a precision of $\approx\unit[0.5]{ms}$, which corresponds to a constraint on the neutron star radius of $\approx\unit[0.7]{km}$ (90\% credibility). This radius constraint is wider than the constraint obtained by measuring the tidal deformability of the neutron star of the same system during the inspiral. Moreover, the neutron star radius is likely to be more tightly constrained using binary neutron star mergers. While NSBH mergers are important for the information they provide about stellar and binary astrophysics, they are unlikely to provide insights into nuclear physics beyond what we will already know from binary neutron star mergers. 
 
\end{abstract}

\section{Introduction}
\label{sec:intro}
Neutron star-black hole (NSBH) mergers make up approximately two of the $\simeq$ 90 gravitational-wave events observed by the LIGO--Virgo--KAGRA (LVK) collaboration \citep{GWTC2, gwtc3, O3b_population, nsbh_discovery_2021}\footnote{We refer here to the two unambiguous NSBH detections described in \cite{nsbh_discovery_2021}. However up to four NSBH events may be included in the current catalog of events \citep{gwtc3}.}. While the first binary neutron star detection was a multi-messenger discovery \citep{BNS_discovery_170817}, neither of the NSBH events observed so far have been associated with a known electromagnetic counterpart. Observations of both binary neutron stars and NSBH systems can provide clues about the neutron star equation of state \citep[e.g.,][]{Lindblom_1992L,kochanke_1992, bejger_2005, Flanagan_2008, Read_2009, Duez_2010, Pannarale_2011, Lattimer_2016, Oertel_2017, 170817_eos_2018}. 
If the black hole is sufficiently light or spinning rapidly, the neutron star may tidally disrupt. Such a disruption could provide a progenitor for a gamma ray burst \citep[e.g.,][]{Mochkovitch_1993,Janka_1999,NAKAR_2007}, a kilonova \citep[e.g.,][]{Li_1998, Metzger_2010, kawaguchi_2016}, and the creation of $r$-process elements \citep[e.g.,][]{Freiburghaus_1999}. 

When the neutron star in an NSBH disrupts, the gravitational-waveform cuts off abruptly. The frequency at which this occurs depends on the binary system parameters, including the equation of state of the neutron star \citep{vallisneri_2000, Ferrari_2010}. The relationship between the neutron star equation of state and tidal disruption has been explored in numerical-relativity studies \citep[e.g.,][]{lackey_2014,foucart_2013_eos, pannarale_2015_amplitude}. However, most NSBH mergers are probably not associated with tidal disruption. A low mass ratio $q = m_2/m_1 \lesssim 1/6$, a high prograde black-hole spin ($\chi_\text{BH} \gtrsim 0.5$), and a large neutron-star radius are likely necessary ingredients for disruption \citep{Kyutoku_2011, Foucart_2013, Hannam_2013}. Highly spinning black holes seem difficult to produce in most binary stellar evolution scenarios, since first-born black holes are thought to be non-spinning \citep[e.g.,][]{Fuller2019}, although spins may be induced through mass-transfer or Wolf-Rayet stellar winds \citep[e.g.,][]{Steinle_2023}. This may suggest that some or most disrupting NSBH come from binaries that have undergone mass ratio reversal during their evolution; the neutron star is born first and the secondary star, which will go on to become a black hole, is spun up through tidal interactions \citep{Qin_2018, Chattopadhyay_2022, Hu_2022}. This scenario could account for up to 20\% of the NSBH population \citep{broekgaarden_2021}. 

We expect to detect gravitational waves from 1 -- 180 NSBH mergers per year \citep{Abadie_2010, Baibhav_2019, broekgaarden_2021} at LVK design sensitivity \citep{adv_ligo_2015, AdvancedVirgo, 2020_Kagra}. However, the fraction of systems that will undergo tidal disruption will likely only comprise around 10\% of this population \citep{Kumar_2017, Zappa_2019, Zhu_2021, Fragione_2021, Biscoveanu_2023}. Even if disrupting binaries are a small minority of NSBH systems, future detectors such as Cosmic Explorer \citep{CE} and Einstein Telescope \citep{ET} will allow us to observe a population of potentially disruptive binaries \citep[e.g.,][]{gupta_2023}. 

There are several NSBH waveform approximants available that model the disruption of the neutron star. For example, \cite{lackey_2014, Thompson_2020, Matas_2020, Gonzalez_2022} all present waveform models tuned to numerical-relativity simulations of NSBH binaries. They include prescriptions to damp the gravitational waveform at the onset of tidal disruption.
In these models, the disruption time is determined by the neutron star equation of state. 
The mass and spin of the remnant black holes are informed by the models developed by \cite{Pannarale_2013,Pannarale_2014} and later \cite{Zappa_2019}. 
These models classify the mergers according to the descriptions of \cite{pannarale_2015_amplitude,pannarale_2015} as disruptive, non-disruptive, and mildly disruptive to further improve the waveform accuracy when compared to numerical relativity.

In this Letter we show that the disruption of a nearby NSBH system can be measured in gravitational-wave data. We show that this can be used to constrain the neutron star radius, though this constraint is wider than the constraint obtained from measuring the neutron star tidal deformability parameter during the inspiral. The remainder of this Letter is organised as follows. In Section \ref{sec:model} we describe a phenomenological model for the gravitational-wave signal from a disrupting NSBH. In Section \ref{sec:simulations} we describe our analysis of a simulated NSBH signal. In Section \ref{sec:Result} we show that, while we can measure the disruption time of the system to within $\lesssim\unit[1]{ms}$, this measurement does not translate to a superior constraint on the neutron star radius compared to measuring the tidal deformability during the inspiral. We show that the tidal deformability measurement  provides a more precise constraint on the radius and therefore the equation of state. Moreover, NSBH disruptions provide less information about the neutron-star equation of state than what we will learn from measurements of tidal effects with merging binary neutron stars. \\

\section{Disrupting NSBH waveform model}\label{sec:model}
For the waveform models mentioned above, the disruption time is determined by the binary parameters and the assumed neutron star equation of state. In this analysis, we seek to determine the extent to which the disruption can be ``seen,'' and how much information is provided by the observation of the disruption versus the observation of tidal effects. To this end, we treat the disruption time as an independent parameter, even though in reality, it is determined by the other binary parameters and the equation of state. This allows us to answer questions like: ``with what precision can we measure the disruption time?'' which we could not ask if the disruption time was already determined by the measurement of binary parameters obtained from the inspiral.

\begin{figure}
    \centering
    \includegraphics[scale=0.35]{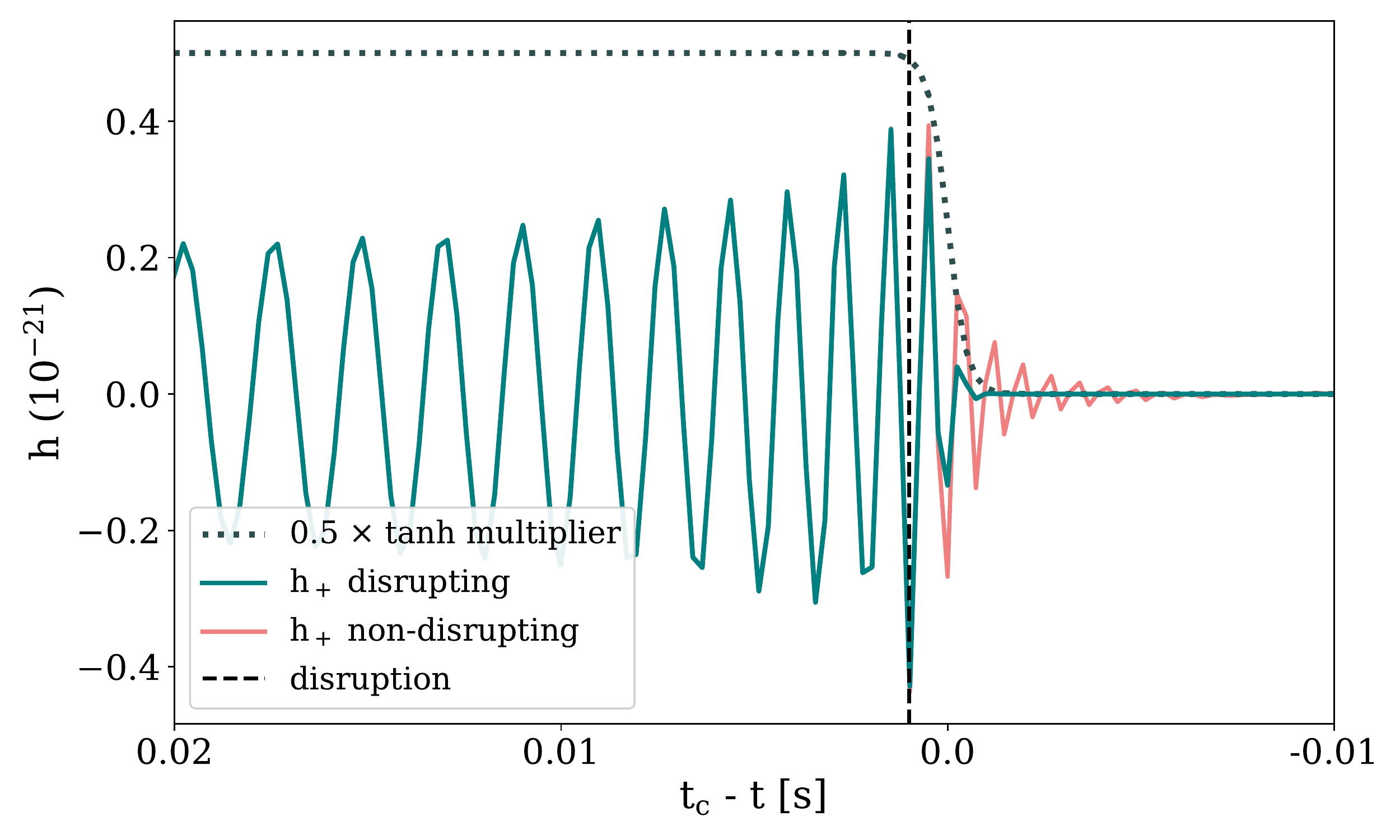}
    \caption{Phenomenological model for the  gravitational waveform of a disrupting neutron star-black hole merger (teal curve).
    The disruption, which here begins $ \lesssim \unit[1]{ms}$ before merger, is indicated by the dashed vertical line.
    The dotted curve shows the $\tanh$ function (Equation \ref{eq:tanh_func}) that we use to cut off the gravitational-wave signal that would be present if the neutron star did not disrupt. 
    In pink we show the the original \texttt{IMRPhenomPv2\_NRTidal} waveform (with no disruption).
    The parameters for this binary are provided in Table~\ref{tab:params}.}
    \label{fig:waveform}
\end{figure}

We use the binary neutron star waveform \texttt{IMRPhenomPv2\_NRTidal} \citep{Dietrich_2019}, denoted $h_\text{BNS}(t)$, as a starting point to construct a phenomenological model of a disrupting NSBH system. We also treat the neutron-star tidal deformability as a free parameter. We also sample over the tidal deformability of the primary object, which we model as a black hole with tidal deformability $\Lambda_{1}=0$. Using these assumptions, we obtain a preliminary \texttt{IMRPhenomPv2\_NRTidal} waveform. We multiply this waveform by a hyperbolic tangent ``window'' function $w(t|\tau_d, \Delta t)$, inspired by the amplitude corrections employed by \cite{lackey_2014,Thompson_2020,Matas_2020}, which simulates a disruption by prematurely terminating the waveform:
\begin{equation}
    h(t) = h_\text{BNS}(t) \, w(t|\tau_d, \Delta t) , 
    \label{eq:signal_model}
\end{equation}
where
\begin{equation}
   w(t|\tau_d, \Delta t) =  \frac{1}{2} + \frac{1}{2}\tanh\Big(\frac{2(\tau_d + \Delta t - t)}{\Delta t}\Big) .
   \label{eq:tanh_func}
\end{equation}
Here, $\tau_d$ is a free parameter describing the time of disruption relative to the merger time of the original \texttt{IMRPhenomPv2\_NRTidal} waveform.
Meanwhile, $2\Delta t$ is a free parameter describing the time interval over which the strain goes to zero. 
Figure \ref{fig:waveform} shows an example tidal-disruption waveform (teal) alongside the original \texttt{IMRPhenomPv2\_NRTidal} waveform (pink).

\section{Analysis of a simulated NSBH signal}
\label{sec:simulations}
We inject a tidally disrupting NSBH signal into the simulated Gaussian noise of two \unit[40]{km} Cosmic Explorer detectors \citep{CE_psd_2017} located at the sites of LIGO Hanford and LIGO Livingston.\footnote{We use the Figure 1 Cosmic Explorer power spectral density noise curve from \cite{CE_psd_2017} which is available as supplementary data.}. We optimistically choose binary parameters (shown in Table \ref{tab:params}) to achieve a tidal disruption at approximately the lowest realistic frequency given our understanding of compact objects: $\approx\unit[1600]{Hz}$. In order to achieve this low frequency, the black hole has a mass of only $3.6 M_\odot$ with a high prograde spin of $\chi_1 = 0.9$. We assume aligned spins in this study. The $1.2 M_\odot$ neutron star has a dimensionless tidal deformability of $\Lambda_2 = 960$ corresponding to a radius of $\unit[12]{km}$. This is a relatively stiff choice of equation of state given that \cite{170817_eos_2018} constrains $\Lambda_{1.4} = 190^{+390}_{-120} $ and $ \unit[R_{1.4}=11.9^{+1.4}_{-1.4}]{km}$ with the binary neutron star merger GW170817 but is within the constraints calculated using NICER mass and radius measurements of pulsars \citep{Miller_2021,Legred_2021}. Appendix \ref{sec:param_choice} provides further details on our choice of parameters. We optimistically place the merger at a distance of $\unit[150]{Mpc}$, which is closer than both of the NSBH events observed by the LVK \citep{nsbh_discovery_2021}. This corresponds to an optimal network signal-to-noise ratio (SNR) of $\approx 335 $ in our Cosmic Explorer network.
We use \texttt{Bilby} \citep{Ashton_2019_bilby, Romero_Shaw_bilby} and \texttt{Parallel\_Bilby} \citep{pbilby_paper} to perform Bayesian inference on a simulated NSBH system. 
In order to reduce the computational challenge of this calculation, and because we are primarily interested in the disruption and tides of the system, we begin our analysis at \unit[100]{Hz} and only analyse the final \unit[12]{s}  of the signal up to a frequency of \unit[2048]{Hz}. 
We fix some parameters to their injected values, which Cosmic Explorer is predicted to be able to measure extremely accurately during the inspiral for such a nearby source: $m_1$, $m_2$, RA, Dec \citep[e.g.,][]{Vitale_2017,Vitale_2018, Smith_2021, Borhanian_2022, Iacovelli_2022}. 
\cite{Iacovelli_2022} forecast the precision with which different parameters can be measured for NSBH in a three-detector third-generation network. They predict that at SNR 300: the chirp mass can be measured with a fractional uncertainty of $\approx 10^{-5}$, the symmetric mass ratio can be measured with a precision of $\approx 10^{-4}$ and the sky position can be measured with a precision of $\approx 0.1 \text{deg}^2$.
We fix the duration of the tidal disruption $\Delta t = \unit[2]{ms}$, based on numerical relativity studies that show the gravitational-wave amplitude decaying on the  timescale of $\approx$ \unit[few]{ms} \citep[e.g.,][]{Yamamoto_2008, shibata_2009, Chaurasia_2021}. We sample over the remaining parameters 
\begin{align}
    \Theta = \{\chi_1, \chi_2, \phi_c, \theta_\text{JN}, \psi, \Lambda_1, \Lambda_2, t_\text{geo}, \tau_d \}.
\end{align}
Table \ref{tab:params} shows the injection parameters and priors used for our simulation. We use the nested sampler \texttt{dynesty} \citep{dynesty} to obtain posterior samples for $\Theta$.

\begin{table}
\caption{Source parameters of the simulated waveform. The disruption time $\tau_d$ is parameterised relative to the coalescence time $t_c$ of a non-disrupting system. The uniform prior listed for luminosity distance refers to uniform in co-moving volume.}
\begin{center}
\begin{tabular*}{\columnwidth}{ c c c c }
 \hline
 Parameter & Abbreviation & Value & Prior  \\ [0.5ex] 
 \hline
 masses & $m_1$, $m_2$ & 3.6, 1.2 M$_\odot$ & fixed \\ 
 
 spin parameters & $\chi_1$,$\chi_2$ & 0.9, 0.0 & \shortstack{U(-1,1), \\ U(-0.5,0.5)} \\
 
 inclination & $\theta_\text{JN}$ & 0.1 & Sine(0,$\pi$) \\
 
 phase & $\phi_c$ & 1.3 & U(0,2$\pi$)\\

 polarisation & $\psi$ & 2.7 & U(0, $\pi$) \\
 
 luminosity distance & D$_L$ & 150 Mpc & U(10,200) \\
 
 right ascension & RA & $\unit[1.375]{hr}$ & fixed\\

 declination & Dec & $-1.21^\circ$ & fixed\\

 geocentric time & t$_\text{geo}$ & \shortstack{11262-\\59642 s} & \shortstack{U( t$_\text{geo}$-0.3,\\ t$_\text{geo}$+0.6)} \\

 tidal deformability & $\Lambda_1$, $\Lambda_2$ & 0, 960 & \shortstack{U(0,50), \\ U(0,2500)} \\

 disruption time & $t_c - \tau_d$ & 0.4 ms & U(-1,5)\\

 disruption interval & $\Delta t$ & 2 ms & fixed\\
 \hline
\end{tabular*}
\label{tab:params}
\end{center}
\end{table}

There is a one-to-one mapping between disruption time $\tau_d$ and neutron-star radius $R_\text{NS}$.
First, we map the disruption time to disruption frequency using the functional definition of the \texttt{IMRPhenomPv2\_NRTidal} phase as implemented in \texttt{LalSimulation} \citep{ lalsuite, garcia_2020}. 
The derivative of the phase as a function of time gives the gravitational-wave frequency as a function of time.
Next, we solve for the neutron-star compactness using the ``EOS independent'' fitting formula from \cite{pannarale_2015}, which relates the compactness to the disruption frequency in dimensionless units:
\begin{align}\label{eq:f_cut}
    f_\text{cut} = & \sum^3_{i,j,k=0} f_{ijk} \, \mathcal{C}^i \, q^{-j} \, \upchi^k
\end{align}
where 
\begin{align}
    i+j+k \leq & 3 .
\end{align}
Here, $\mathcal{C}$ is the neutron-star compactness, $q$ is the mass ratio of the binary ($m_2/m_1$) and $\upchi$ is the dimensionless spin of the black hole. 
The fitting coefficients $f_{ijk}$ are listed in \cite{pannarale_2015}. We ensure our chosen system is a disrupting binary in this prescription using Equation 2 of \cite{pannarale_2015}, which provides the minimum $q$ for a given NSBH system. Equation \ref{eq:f_cut} is plotted as a function of mass ratio in Figure \ref{fig:f_Q_plot}. 
The neutron star radius is $R_\text{NS} = M_\text{NS} / \mathcal{C}$. 

Of course, the neutron star radius can also be inferred using the tidal deformability of the neutron star, which can be measured precisely by Cosmic Explorer during the inspiral. 
We convert measurements of deformability $\Lambda_2$ to  compactness $\mathcal{C}$ using the universal relations between the quadrupolar dimensionless tidal deformability and the neutron star compactness \citep[e.g.,][]{maselli_2013, Godzieba_21, pradhan_22}. The latter two studies fit a 6th-order polynomial for the $\mathcal{C} - \Lambda_2$ relation:
\begin{equation}
    \mathcal{C} = \sum^6_{k=0}a_k(\mathrm{ln} \ \Lambda_2)^k
\label{eq:LoveC}
\end{equation}
We use the relation and fitting coefficients developed by \cite{pradhan_22}. 

\section{Results and Discussion}\label{sec:Result}
\begin{figure}
    \centering
    \includegraphics[width=\columnwidth]{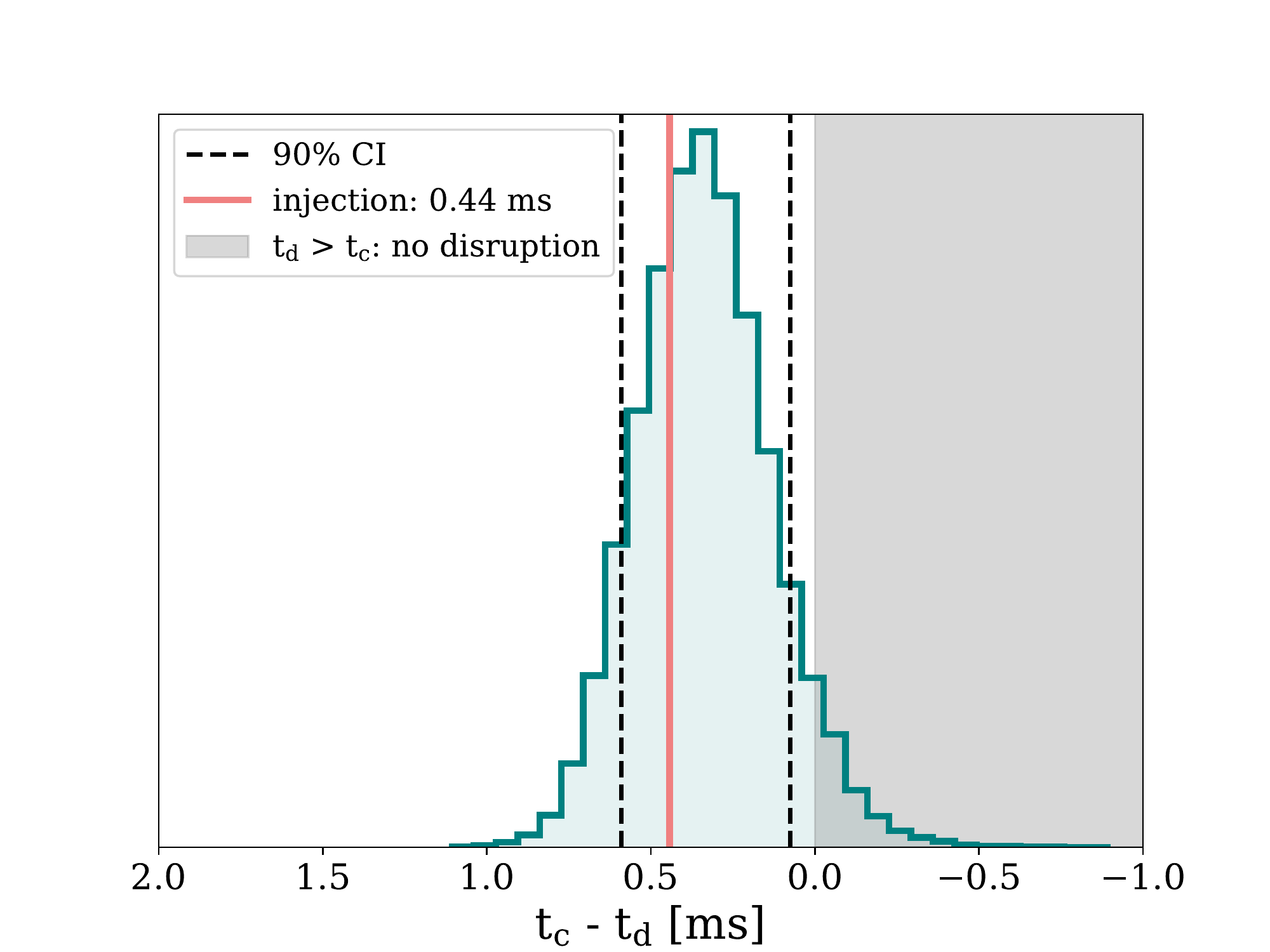}
    \caption{Posterior distribution for the inferred disruption time. The grey region shows the region of parameter space corresponding to there being no disruption as the sampled disruption time takes place after the binary coalescence. This is ruled out at over 90 \% confidence in this simulation. The disruption time is constrained to within \unit[0.5]{ms} (90 \% credibility). }
    \label{fig:time_hist}
\end{figure}

\begin{figure}
    \centering
    \includegraphics[width=\columnwidth]{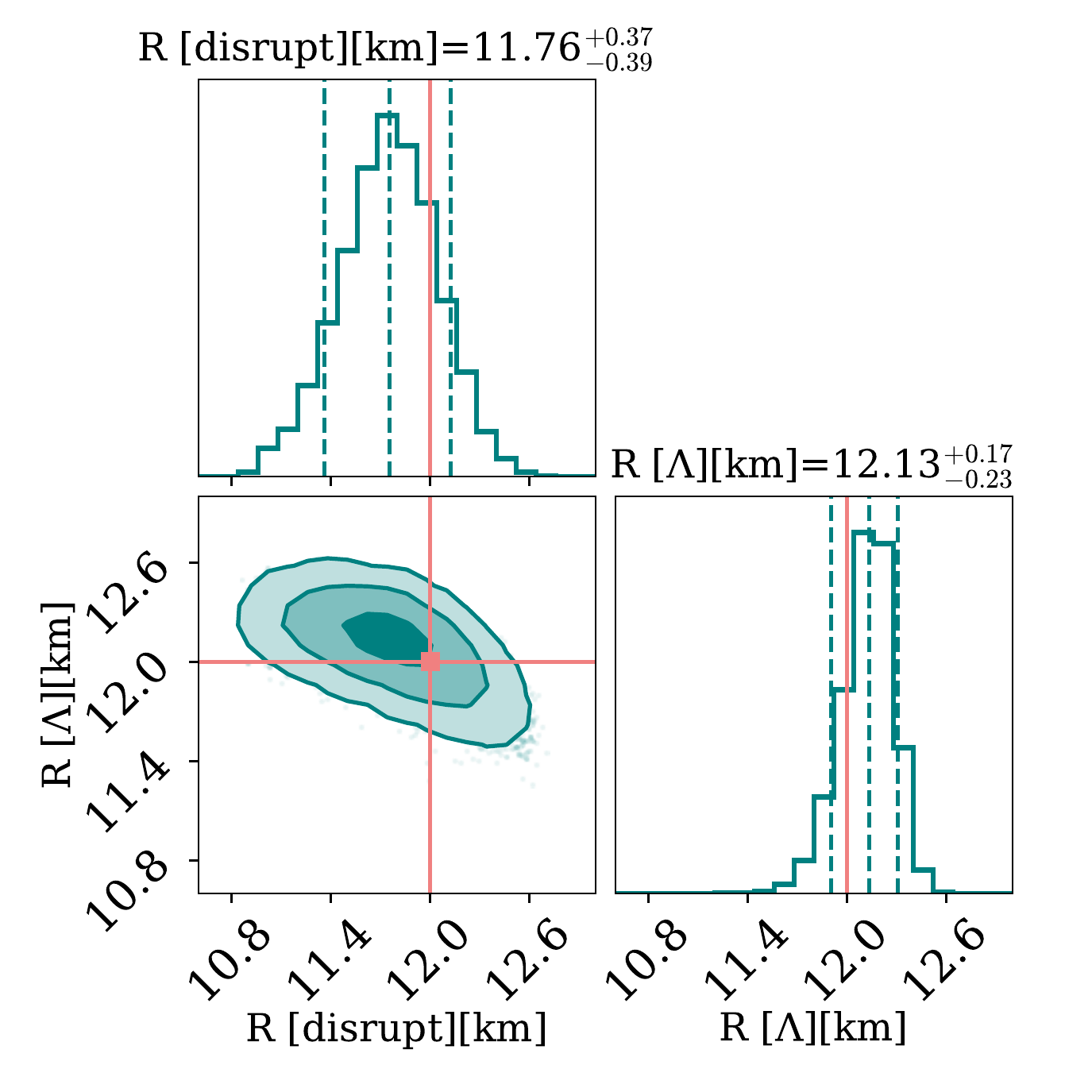}
    \caption{Posterior distribution for the neutron star radius recovered using the tidal disruption and the tidal deformability. The different shades show the one-, two- and three-sigma intervals while the 90\% confidence intervals are shown in the 1d histograms. While both results are consistent with the simulated radius of \unit[12]{km}, the radius is better constrained using the tidal deformability compared with using the tidal disruption. The radius inferred from the tidal disruption is accurate to $\approx$ 7\% while the radius inferred through the tidal deformability is accurate to $\approx$ 3\% at 90\% credibility.}
    \label{fig:radius_corner}
\end{figure}

Figure \ref{fig:radius_corner} shows the marginalised distribution for the neutron star radius, calculated from recovering the tidal disruption (R[disrupt]) and the tidal deformability (R[$\Lambda$]) by performing parameter estimation on a simulated Cosmic Explorer signal at $\unit[150]{Mpc}$. The full posterior distribution is displayed in Figure \ref{fig:corner}. We are able to measure the time of the neutron star disruption to a sensitivity of $\approx \unit[0.5]{ms}$ (90\% credibility).\footnote{This estimate is actually slightly optimistic because we have ignored the effect of the free spectral range (FSR), which causes the detector response to fall off at high frequencies. For Cosmic Explorer, the FSR is $\approx\unit[3.75]{kHz}$ and so the gravitational-wave strain will be attenuated by a factor of $\approx 20\%$ at $\unit[1600]{Hz}$ when the neutron star disrupts; see \cite{Essick}.}
We are also able to rule out a non-disrupting system, which we define as a system with a disruption occurring after the coalescence time of the simulation, at just over 90\% confidence. This is shown in Figure \ref{fig:time_hist}. We remove samples with a disruption time that takes place after the coalescence time before continuing the analysis. This does not significantly change our inferred results or uncertainties but removes a low-radius tail from the posterior distribution for the radius. This measurement results in the disruption frequency being measured to within $\simeq$ \unit[100]{Hz} (90\% credibility), which translates into a radius precision of $\approx 0.7$ km. The results are consistent with the implied `injection' of \unit[12]{km}. 

Figure \ref{fig:radius_corner} also shows the recovery of the tidal deformability parameters and associated measurement of the neutron star radius. This measurement recovers the radius to a higher precision than using the tidal disruption at \unit[0.4]{km} (90 \% credibility). Since the recovery of both the tidal disruption and tidal deformability improves with increasing SNR \footnote{Here we assume the waveform mismatch from numerical relativity is not higher than the statistical uncertainties, which may not be the case with current waveforms at Cosmic Explorer sensitivity, highlighting the need for waveforms calibrated to a larger set of numerical relativity simulations \citep[e.g.,][]{Huang_2021}}, we hypothesise that the radius recovered with the tidal deformability will always outperform the tidal disruption measurement. Measuring the neutron star tidal deformability with a binary neutron star merger will likely be more informative than an NSBH, as the tidal deformability of a binary neutron star at 100 Mpc in Cosmic Explorer is projected to be constrained to 3 \% \citep[e.g.,][]{Martynov_2019}. Constraining the neutron star equation of state using electromagnetic detections of NSBH is likely to be similarly less informative than the information gained using binary neutron stars \citep{Biscoveanu_2023}, especially given that NSBH electromagnetic counterparts are an unlikely outcome of most NSBH mergers \citep[e.g.,][]{Fragione_2021}. We repeat the analysis with a more rapid tidal disruption: $\Delta t = \unit[0.6]{ms} \ (\approx 1/\mathrm{f_d})$, since a sharper cutoff of the gravitational waves could be easier to resolve in parameter estimation. In this simulation, the 90 \% credible intervals for the radius shrink by $\approx$ 25 \%, but are still wider than those obtained from the neutron star tidal deformability. However, the confidence in measuring the tidal disruption through the disruption time increases to $\gtrsim$ 99 \%. Hence, we find abrupt tidal disruptions are advantageous for confidently measuring the disruption in gravitational-wave data but still do not surpass the constraints from the tidal deformability. 

Despite not directly improving our prospects for measuring the neutron star equation of state, measuring the neutron star radius from both the tidal disruption and tidal deformability will provide a valuable consistency check for our understanding of nuclear matter. If the measurements from the tidal disruption and tidal deformability are inconsistent this could indicate shortcomings with our understanding of the behaviour of matter inside neutron stars. Such a measurement could provide hints of new physics that could not be discovered with tidal deformability alone.

The mass of the ejecta expelled by the disruption can be indirectly inferred using the neutron star compactness and the radius of the innermost stable orbit \citep[e.g.,][]{foucart_2012, kawaguchi_2016, Kruger_2020}. We find the compactness is better constrained using the tidal deformability of the neutron star. The innermost stable orbital radius can be calculated using the inferred mass and spin of the black hole \citep{Bardeen_1972}. Hence, provided that a disruption has occurred, the mass of the ejecta can be predicted without directly measuring the disruption in the gravitational-wave data. 

The disruption ejecta are likely dispersed anisotropically \citep[e.g.,][]{kyutoku_2013,kawaguchi_2016}. In theory, measuring the gravitational-wave \textit{phase} of the tidal disruption could be used to estimate the direction of the relativistic ejecta, which could help us better understand kilonovae physics. The ``phase of disruption'' can be estimated as:
\begin{align}
\phi_d = \phi_0 + 2\pi f(t=\tau_d) \tau_d ,
\end{align}
where $\phi_0$ is some reference phase determined by the inspiral and $f(t)$ is the frequency evolution of the gravitational-wave signal.
In order to obtain an interesting estimate of $\phi_d$, it would be necessary to measure $\tau_d$ with a precision that is small compared to the inverse disruption frequency:
\begin{align}
    \sigma_{\tau_d} \ll 1/f_d .
\end{align}
We estimate an additional factor of 5--10 in SNR is required to resolve $\phi_d$, which would require an unrealistically close source. 

The relations we use to infer the neutron star compactness in this work (Eq. \ref{eq:f_cut}, Eq. \ref{eq:LoveC}) are limited by fitting to a finite number of and type of equation of state. While both relations have maximum errors less than 5\% for the range of equations of state tested (and are better than 2\% accurate for 90\% of the parameter space), if neutron stars contain exotic physics, such as phase transitions in the core, then these relations may not properly map between the observable parameters and the compactness. This could introduce new uncertainty into our analysis. \cite{Raithel_2022, Raithel_2022b} present ``tidal deformability doppleg\"angers'', alternative equations of state that differ by up to $\simeq$ \unit[0.5]{km}, despite having almost identical curves in mass-tidal deformability space. This means that our posterior on the radius calculated from the tidal deformability may not be as well-constrained as it appears. However, since our measurement of the radius from the tidal disruption is also based on a fit to numerical relativity simulations, our statements about the relative usefulness of neutron star-black hole mergers are likely unaffected. 

\section{Summary and Conclusions}\label{sec:conclusion}
In this Letter, we use a phenomenological waveform model to measure the tidal disruption of a neutron star in an NSBH merger. We show that the tidal disruption can be observed in gravitational waves for favourable systems with a Cosmic Explorer network. We measure the time of the tidal disruption to $\simeq$ \unit[0.5]{ms}. While this does not allow us to place tighter constraints on the neutron star radius than we obtain with the tides, this method provides us with an independent measure of the neutron star compactness and may help reveal the existence of exotic phase transitions within neutron stars.

We explore other areas of interest related to the neutron disruption and equation of state, such as the mass of the ejected matter and the possibility of measuring phase transitions in neutron stars.
However we ultimately find that the tidal disruption information may not be very helpful in solving these problems. We suggest that while NSBH systems are important for understanding stellar and binary evolution, they do not provide crucial information for decoding the nuclear equation of state. Future studies should further consider the implications of phase transitions on the equation of state measurable with gravitational waves.

\begin{acknowledgments}
We thank the referee for their helpful suggestions which improved this manuscript. We thank the Caltech gravitational-wave group, especially Isaac Legred for interesting and helpful discussions about this work. This work is supported through Australian Research Council (ARC)  Centre of Excellence CE170100004, Discovery Projects DP220101610 and DP230103088, and LIEF Project LE210100002.  T. A. C. receives support from the Australian Government Research Training Program. The authors are grateful for for computational resources provided by the LIGO Laboratory computing cluster at California Institute of Technology supported by National Science Foundation Grants PHY-0757058 and PHY-0823459, and the OzSTAR Australian national facility at Swinburne University of Technology.
\end{acknowledgments}

\appendix
\section{Choice of simulation parameters}
\label{sec:param_choice}
\begin{figure}
    \centering
    \includegraphics[scale=0.6]{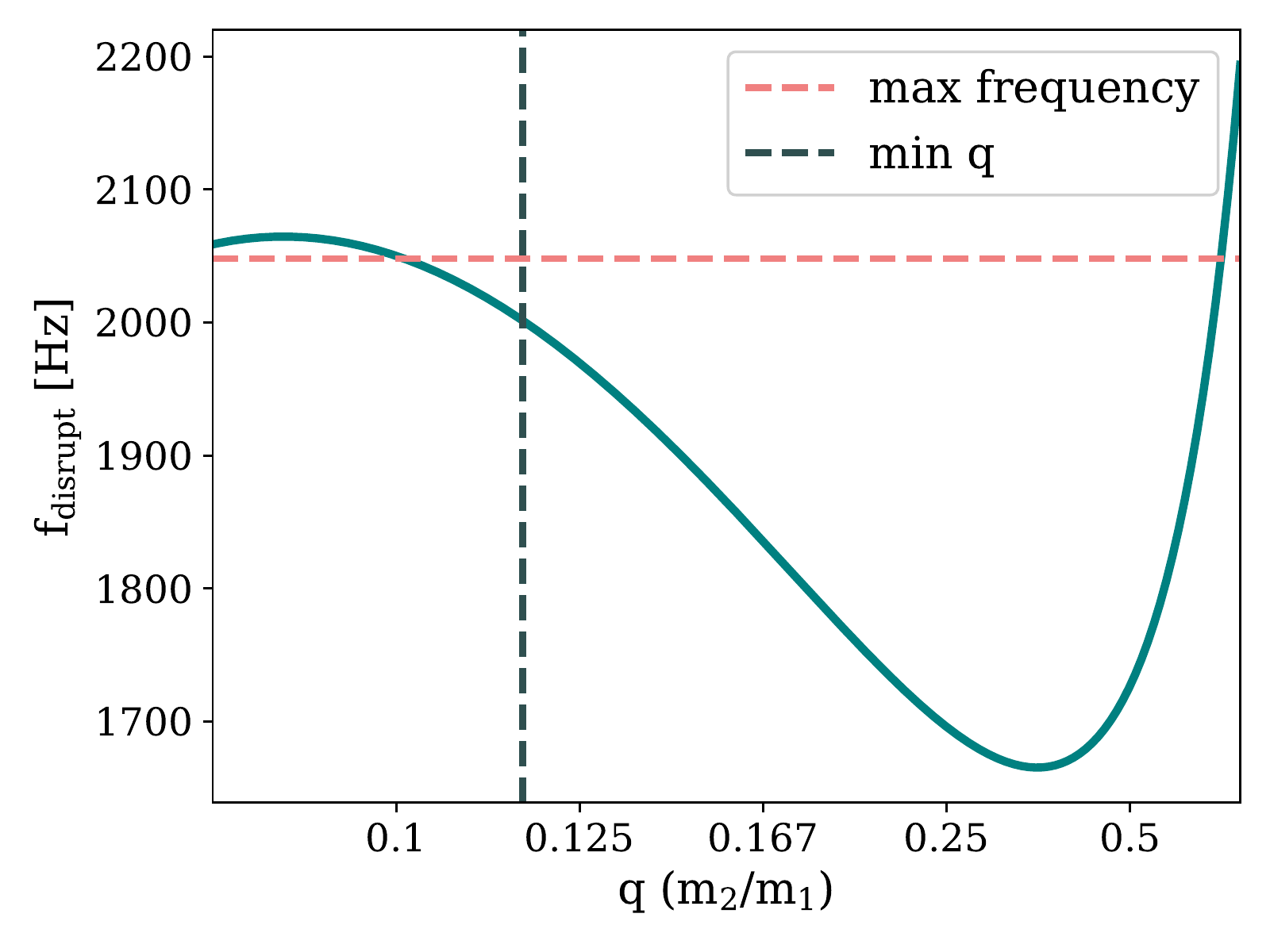}
    \caption{The disruption frequency calculated from Equation \ref{eq:f_cut} as a function of mass ratio for an NSBH system for an NSBH with $m_\text{NS} = \unit[1.2]{M_\odot}$, $R_\text{NS}=\unit[12]{km}$ and $\chi_\text{BH} = 0.9$. This system produced the lowest disruption frequency for any ``reasonable" choice of parameters. Increasing the neutron star mass or decreasing the black hole spin or neutron star radius primarily shifts the curve upwards in frequency-space. The optimal disruption frequency is $\approx$ \unit[1660]{Hz} at a mass ratio of $\approx 1/3$ . The dotted lines show the maximum frequency of the waveforms we use in parameter estimation and the minimum mass ratio for which the system is disruptive. Hence, only systems in the bottom right quadrant of the plot are both disruptive and have low enough frequencies to be potentially resolvable in our analysis.}
    \label{fig:f_Q_plot}
\end{figure}

\begin{figure}
    \centering
    \includegraphics[scale=0.6]{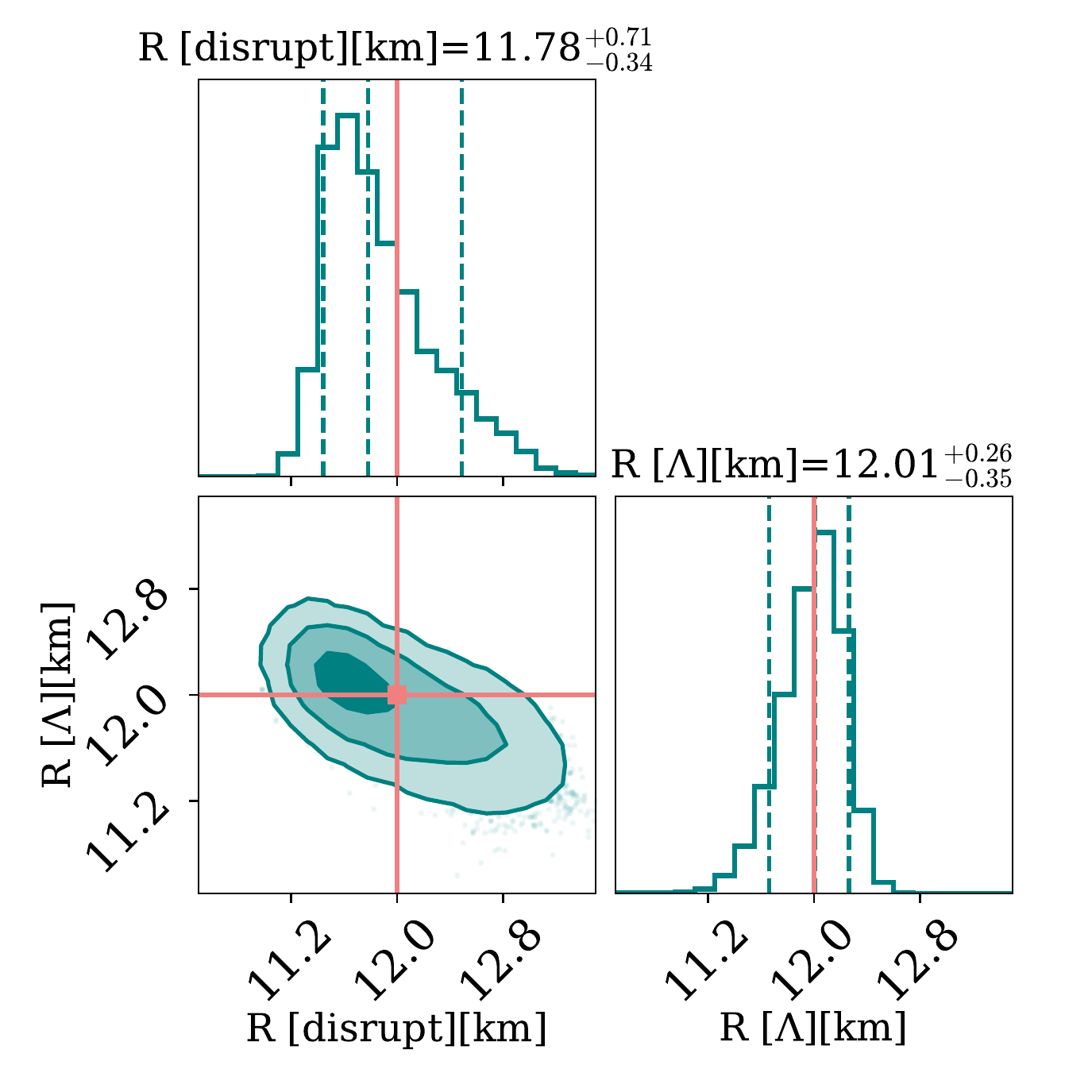}
    \caption{Posterior distribution for the neutron star radius recovered using the tidal disruption and the tidal deformability for a system with a mass ratio of 0.2. The different shades show the one-, two- and three-sigma intervals while the 90\% confidence intervals are shown in the 1d histograms. Like the $q=1/3$ system, the radius is better constrained using the tidal deformability compared with using the tidal disruption. The radius inferred from the tidal disruption is accurate to $\approx$ 9\% while the radius inferred through the tidal deformability is accurate to $\approx$ 5\% at 90\% credibility.}
    \label{fig:radius_cornerQ5}
\end{figure}

We choose our system parameters such that the disruption frequency is minimised for a realistic disrupting binary. Figure \ref{fig:f_Q_plot} shows the disruption frequency as a function of mass ratio for an NSBH with $m_\text{NS} = \unit[1.2]{M_\odot}$, $R_\text{NS}=\unit[12]{km}$ and $\chi_\text{BH} = 0.9$. 
We find that a mass ratio of $q \approx 1/3$ provides the optimal conditions to resolve the neutron star disruption given a fixed neutron-star mass of $m_\text{NS}=1.2 M_\odot$, a black-hole spin of $\chi_\text{BH}=0.9$, and a neutron-star radius of $R_\text{NS}=\unit[12]{km}$. 
We verify this by testing a system with the same parameters $(m_\text{NS}, \chi_\text{BH}, R_\text{NS})$ but a mass ratio of 0.2. In other words, we increase the black hole mass while keeping all other parameters fixed.
On the one hand, increasing the black-hole mass increases the overall strain amplitude, which increases the optimal SNR.
On the other hand, it increases the disruption frequency, pushing it into the shot noise where it is difficult to resolve \citep[e.g.,][]{Brown_2022}.\footnote{It also causes the tidal deformability to be less resolvable because the post-Newtonian expression that introduces tides includes the mass ratio \citep[e.g.,][]{lackey_2014,Dietrich_2019,Coupechoux_2022}.}
The loss of SNR from the increased disruption frequency is the more important effect. Systems with mass ratios of 0.2 or less are likely more abundant in nature than those with larger mass ratios \citep[e.g.,][]{Biscoveanu_2023}. 
The neutron star radius posterior distributions for this run are shown in Figure \ref{fig:radius_cornerQ5}.

\section{Full posterior distribution}
\label{sec:full_posterior}
\begin{figure*}
    \centering
    \includegraphics[width=\columnwidth]{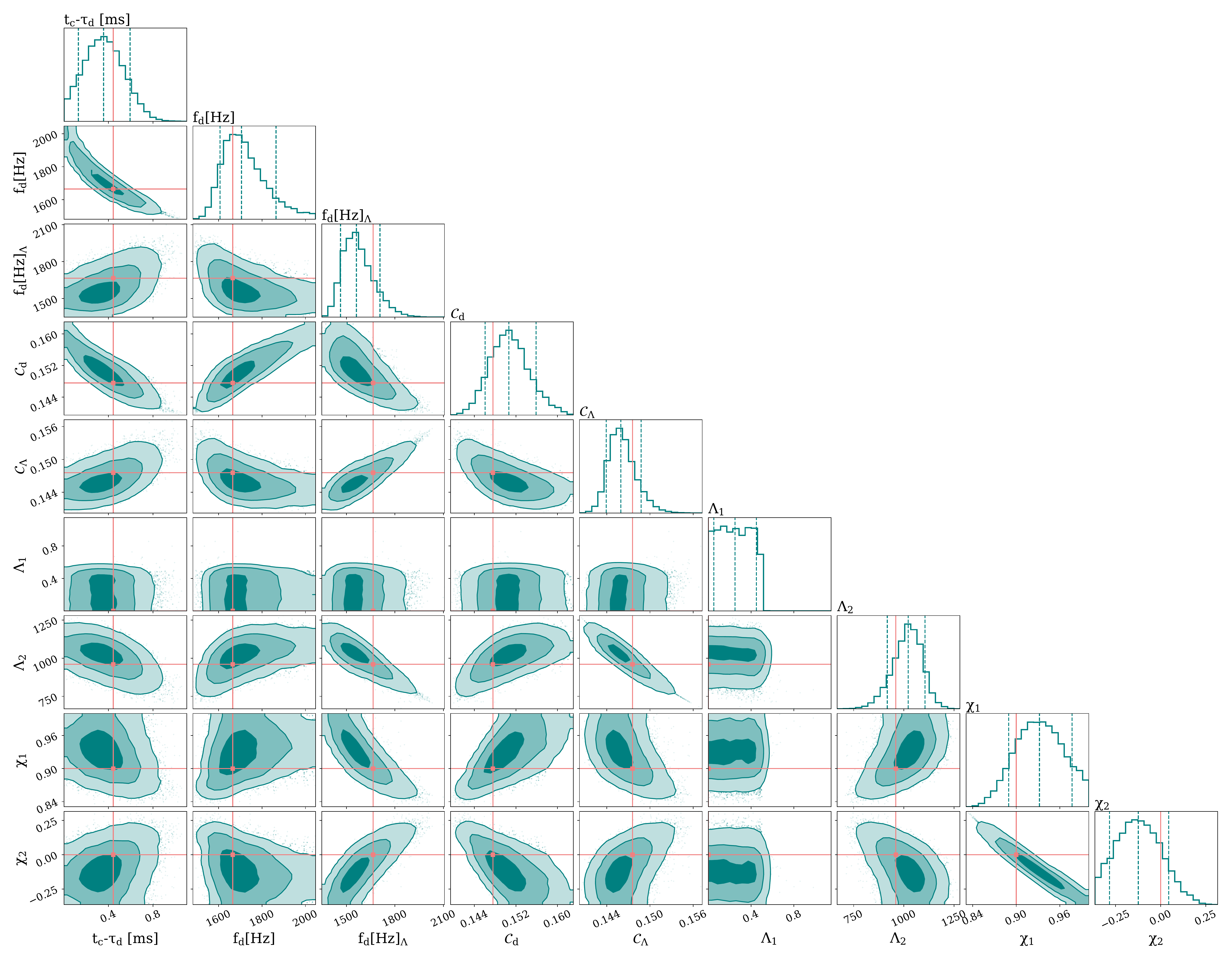}
    \caption{Posterior distribution for key NSBH parameters measured by a network of Cosmic Explorer observatories.
    The different shades show one-, two-, and three-sigma credible intervals. From left to right, the parameters are: the neutron star disruption time relative to the BNS merger time in seconds $\mathrm{t_c - t_d}$, the disruption gravitational-wave frequency $\mathrm{f_d}$ measured from the disruption time, the gravitational-wave frequency measured indirectly from the neutron star tidal deformability $\mathrm{f_{d [\Lambda]}}$, the neutron star compactness $\mathcal{C}_\text{d}$ and $\mathcal{C}_\Lambda$ measured from the disruption frequency and the tidal deformability respectively, the black hole tidal parameter $\Lambda_1=0$, the neutron star tidal parameter $\Lambda_2$, the black-hole spin magnitude $\mathrm{\chi_1}$ and the neutron-star spin magnitude $\mathrm{\chi_2}$. The posterior distributions for the disruption frequency calculated two ways are comparable, however the posterior distribution for the neutron-star compactness calculated with the tidal deformability is more constraining than when calculated with the disruption information. This is likely because in the kHz regime, small changes in frequency correspond to larger changes in neutron star equation of state than comparable changes in tidal deformability.}
    \label{fig:corner}
\end{figure*}

Figure \ref{fig:corner} shows the posterior distribution of the parameters sampled in our simulation as well as those sampled indirectly in post-processing (disruption frequency and compactness). Parameters labelled with the subscript ``d'' are parameters that were inferred indirectly using the tidal disruption while those labelled with the subscript ``$\Lambda$'' are parameters that were inferred indirectly using the tidal deformability. 

\bibliography{bib}
\bibliographystyle{aasjournal}
\end{document}